\documentstyle[epsfig,12pt]{article}
\makeatletter
\def\Journal#1#2#3#4{{#1} {\bf #2}, #3 (#4)}

\def\NPB{{\em Nucl. Phys.} B}
\def\PLB{{\em Phys. Lett.}  B}

\def\ZPC{{\em Z. Phys.} C}
\newcommand{\rd}{{\mathrm{d}}}
\newcommand{\re}{{\mathrm{e}}}

\newcommand{\rB}{{\mathrm{B}}}

\newcommand{\Wminsq}{{W^2_{\rm min }}}

\newlength{\abstwidth}
\begin{document}
 
\sloppy

\renewcommand{\arraystretch}{1.5}

\pagestyle{empty}

\begin{flushright}
CERN--TH/97--193  \\
ITP-SB-97-48 \\
hep-ph/9708261
\end{flushright}
 
\vspace{\fill}
 
\begin{center}
{\LARGE\bf Model-independent QED corrections to}\\[3mm]
{\LARGE\bf photon structure-function measurements$^{a}$}\\[10mm]
{\Large Eric Laenen}\\[3mm]
{\it Institute for Theoretical Physics, 
     State University of New York at Stony Brook,}\\[1mm]
{\it Stony Brook, NY 11794, USA}\\[1mm]
{E-mail: {\tt eric@insti.physics.sunysb.edu}}\\[2ex]
{\large and} \\[2ex]
{\Large Gerhard A. Schuler$^b$} \\[3mm]
{\it Theory Division, CERN,} \\[1mm]
{\it CH-1211 Geneva 23, Switzerland}\\[1mm]
{ E-mail: {\tt Gerhard.Schuler@cern.ch}}
\end{center}
 
\vspace{\fill}
 
\begin{center}
{\bf Abstract}\\[2ex]
\begin{minipage}{\abstwidth}
We present the first calculation of QED radiative corrections to 
deep-in\-e\-las\-tic electron--photon scattering in terms of those variables 
that are reconstructed in measurements of the photon structure function 
in electron--positron collisions. 
In order to cover the low-$Q^2$ region, we do not invoke 
the QCD-improved parton model but rather express our results 
in terms of the photon structure functions. 
Both analytical and numerical results are given.
\end{minipage}
\end{center}

\vspace{\fill}
\noindent
\rule{60mm}{0.4mm}

\vspace{1mm} \noindent
${}^{a}$ Talk presented at the XIth Workshop on Photon--Photon Collisions, 
Photon '97, Egmond aan Zee, The Netherlands.\\[1mm]
${}^b$ Heisenberg Fellow.

\vspace{10mm}\noindent
CERN--TH/97--193\\
August 1997

\clearpage
\pagestyle{plain}
\setcounter{page}{1} 

QED radiative corrections distort the usual kinematics of deep-inelastic 
scattering (DIS) and hence have to be taken into account 
for precise structure-function measurements. In the case of the 
proton structure function, the Born kinematics (corresponding to 
non-radiative events) of charged lepton--nucleon scattering 
is fully constrained by two measurable variables. 
The arguments of $F_2(x,Q^2)$, Bjorken-$x$ and the squared 
momentum transfer $Q^2$, can directly be determined
from either the scattered lepton or the hadronic system. Alternatively, 
the kinematics can be fixed by two ``mixed'' variables such as
the polar angles of the scattered lepton and the hadronic system. 
Photon radiation affects different variables differently but, in general, 
several variables can be reconstructed experimentally and hence one can has 
experimental cross checks on the size of radiative 
corrections \cite{Spiesberger}.

The situation is more complicated for measurements of the photon
structure function in electron--positron collisions. At given 
$\re^+\re^-$ c.m.\ energy $\sqrt{s} = 2\, E_b$ ($E_b$ is the beam energy), 
three variables are needed 
in order to specify the Born kinematics since the target-photon energy
is not known. (Actually, for precision measurements also the effects 
of the non-zero target mass $P^2$ have to be considered.) 
Moreover, there is just one way to experimentally 
reconstruct three independent variables.

The angle $\theta$ and energy $E$ of the tagged electron give the 
leptonic DIS variables $y_l$ and $Q_l^2$ as follows:
\begin{equation}
  y_l = 1 - \frac{E}{E_b}\, \cos^2\frac{\theta}{2}\ ,
\qquad 
  Q_l^2 = 4\, E\, E_b\, \sin^2\frac{\theta}{2} \ .
\label{tagangle}
\end{equation}
A measurement of the hadronic mass $W_h$ (which involves an unfolding 
of $W_h$ from the visible hadronic energy $W_{\rm{vis}}$) 
yields the ``mixed'' Bjorken-$x$ variable
\begin{equation}
  x_m = \frac{Q_l^2}{Q_l^2 + W_h^2}
\ .
\label{xmdef}
\end{equation}

In general, neither $x_m$ nor $Q_l$ coincide with the actual arguments 
$x_h$ and $Q_h$ of the photon structure function $F_2(x_h,Q_h^2)$, see 
e.g.\ (\ref{varrel}) or Fig.~3 below. Consider 
DIS of electrons on (quasi-real) 
photons in the presence of an additional photon (Fig.~1):
\begin{equation}
 \re(l) + \gamma(p) \rightarrow \re(l') + \gamma(k) + X(p_X) 
\ ,
\label{DISreact}
\end{equation}
and define leptonic and hadronic DIS variables:
\begin{equation} 
\begin{array}{rclrcl}
 q_l & = & l-l' & q_h & = & p_X - p = q_l - k\\
 W_l^2 & = & (p+q_l)^2 \qquad & W_h^2 & = & (p+q_h)^2 = p_X^2\\
 Q_l^2 & = & - q_l^2 &  Q_h^2 & = & - q_h^2\\
 x_l & = & Q_l^2/2p \cdot q_l & 
     x_h & = & Q_h^2/2p \cdot q_h\\
 y_l & = & p \cdot q_l/p \cdot l & 
 y_h & = & p \cdot q_h/p \cdot l
\end{array} 
\label{DISvar}
\end{equation} 
Obviously, leptonic and hadronic variables do not coincide 
($Q_h^2 \neq Q_l^2$, etc.); they agree only for nonradiative
events, i.e. if $k=0$. 

\begin{figure}[tbp]
\begin{center}
{}{\hspace*{\fill}}\psfig{figure=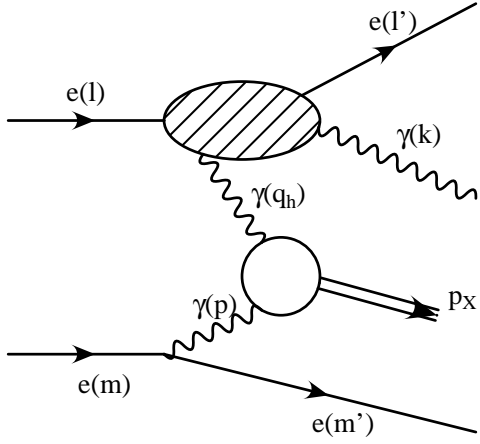,width=0.4\textwidth}%
{}{\hspace*{\fill}}
\caption{Photon bremsstrahlung from the tagged lepton line
in deep-inelastic scattering off an equivalent photon.
\label{fig:one}
}
\end{center}
\end{figure}
In this paper we present the $O(\alpha)$ correction to DIS in terms 
of the experimentally relevant variables $x_m$, $Q_l^2$, and $y_l$:
\begin{equation}
  \frac{\rd^3\sigma}{\rd x_m\, \rd y_l\, \rd Q_l^2}
  = g^\rB(x_m,y_l,Q_l^2,s)  + g^{\rm{corr}}(x_m,y_l,Q_l^2,s)
\ .
\label{totalcross}
\end{equation}
Since the accessable $Q$ values are far below the weak scale, we can 
safely neglect weak corrections apart from the running of the electromagnetic 
coupling $\alpha(Q^2)$. For the case of charged lepton--nucleon scattering 
it is known \cite{Consoli} that QED corrections are very well
approximated by calculations in the leading-log approximation (LLA), 
that is the QED corrections are dominated by photon 
radiation off the tagged-lepton line (Fig.~1). For the case of 
electron--photon scattering, there might be additional, sizeable 
corrections to the untagged electron line. However, we shall show in 
the next paragraph that these are, in fact, small.

The target photon $\gamma(p)$ is part of the flux of equivalent
photons around the non-tagged lepton. To leading order in $\alpha$, 
this flux has a momentum density given by the Weizs\"{a}cker-Williams 
expression $f_{\gamma/\re}(z)$, where $z$ is the longitudinal momentum 
fraction of the target photon with respect to its parent lepton: 
\begin{equation}
  f_{\gamma/\re}(z) = \frac{\alpha}{2\pi} \left\{
  \frac{Y_+(z)}{z}\, \ln \frac{P^2_{max}}{P^2_{min}}
  - 2 m_e^2 z(\frac{1}{P^2_{min}}
 -\frac{1}{P^2_{max}})\right\}
\ .
\label{fww}
\end{equation}
Here we have defined $Y_+(z) = 1 + (1-z)^2$, 
$P^2_{min}  = (z^2 m_e^2)/(1-z)$, $P^2_{max}  = 
(1-z) \left(E_b \theta_{max}\right)^2$, and  
$\theta_{max}$ is the anti-tag\footnote{
i.e. all events in which the parent lepton scatters 
at an angle larger than $\theta_{max}$ are rejected.} 
angle. In the following we put $P^2 \equiv -p^2 = 0$ 
and neglect electron masses everywhere except in (\ref{fww}). 
Moreover we substitute $P^2_{max}$ by $P^2_{max}+P^2_{min}$ so that 
we can easily extend the $z$ range to 1.
QED radiative corrections to this formula follow immediately 
from our previous paper \cite{LaSchu}. The corrections 
can (and must) be resummed and, within the LLA, 
the corrected expression is obtained by replacing 
$Y_+(z)$ in (\ref{fww}) by $Z_+(z,\mu^2)$ defined by
\begin{eqnarray}
  Z_+(z,\mu^2) & = &  Y_+(z)\ \left\{ \exp \left[ \frac{\alpha}{\pi}\ 
       \left( \ln \frac{Q^2}{m_e^2} - 1 \right) \ln(1-z) \right] 
    \right\}
\nonumber\\ & & \quad + \frac{\alpha}{\pi}\ \ln \frac{Q^2}{m_e^2}\, 
  \left\{  z\left(1 - \frac{z}{2}\right) \ln z 
  + z\left(1 - \frac{z}{4} \right) \right\}
\label{resummed2}
\end{eqnarray}
Here $\mu$ is the hard scale of the photon-induced subprocess. 
A numerical evaluation of (\ref{resummed2}) shows that the corrections 
are indeed small, below the few-percent level.

The Born cross section of (\ref{totalcross}) is given by 
\begin{equation}
 g^\rB(x,y,Q^2,s)  = \frac{2 \pi\alpha^2 \,Y_+(y)}{x^2 y^2 s Q^2}
   f_{\gamma/\re}(\frac{Q^2}{xys})\,
   F_2(x,Q^2) \left\{1 + R(x,Q^2,y) \right\}
\ ,
\label{newBorn}
\end{equation}
where
\begin{equation}
 R(x,Q^2,y)  =   \frac{- y^2}{1 +(1-y)^2} \ 
            \frac{F_L(x,Q^2)}{F_2(x,Q^2)} \ ,
\label{Rdef}
\end{equation}
and $F_{2,L}$ are the photon structure functions (we have dropped 
the superscript $\gamma$).

\begin{figure}[tbp]
\begin{center}
{}{\hspace*{\fill}}%
\psfig{figure=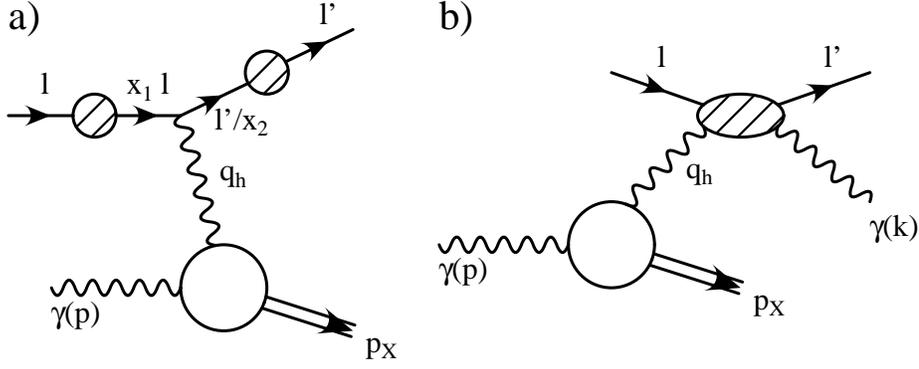,width=0.8\textwidth}%
{}{\hspace*{\fill}}
\caption{a) Initial- and final state bremsstrahlung.
b) Compton contribution.
\label{fig:two}
}
\end{center}
\end{figure}
In the LLA, photon radiation from the tagged lepton line can 
(in a gauge-invariant way) be subdivided into photon bremsstrahlung from 
the initial electron line, the final electron line, and the Compton process,
see Fig.~2. For the cross section relevant for experimental analyses, which is
differential in $x_m$, $Q_l^2$, and $y_l$, there is no 
contribution from final-state radiation as the calorimeter measurement 
combines the electron with the nearby photon(s). 

The initial state radiation correction to the triple differential
cross section in eq.~(\ref{totalcross})
can be written as the following convolution 
\begin{eqnarray}
\lefteqn{
 g^{{\rm ISR}}(x_m, y_l, Q^2_l, s) = 
   \int_0^1 \rd x_1 \, D_{e/e}(x_1,Q_l^2) }
\nonumber\\ & &
  \Big[ \Theta\left( x_1 - x_1^0\right)\, 
     \frac{\hat{x}_m^2}{x_m^2 x_1}\, 
  g^\rB(\hat{x}_m,\hat{y}_l,\hat{Q}^2_l,\hat{s})
       - g^\rB(x_m,y_l,Q_l^2,s) \Big]
\label{bremscross}
\end{eqnarray}
where 
\begin{equation}
  x_1^0 \equiv \frac{x_m\,(1-y_l)\, s + (1-x_m)\, Q_l^2}
  {x_m\, (s - Q_l^2) }
\ ,
\label{x1range}
\end{equation}
and $D_{e/e}(x,Q^2)$ is the structure function for the 
initial-state electron evaluated at the scale given by the squared 
momentum transfer $Q^2 = Q_l^2$:
\begin{equation}
D_{e/e}(x,Q^2)  = \frac{\alpha}{2\pi}\ln\left(\frac{Q^2}{m_e^2}\right)
\, \frac{1+x^2}{1-x}
\ .
\end{equation}
It represents the probability of finding, inside a parent electron, 
an electron with longitudinal momentum fraction $x$. 
The Born cross section 
is written in
terms of the reduced (``hatted'') variables, 
$g^\rB(\hat{x}_m,\hat{y}_l,\hat{Q}^2_l,\hat{s})$. 
The scaling behavior of the
relevant variables is
\begin{eqnarray}
\hat{Q_l^2} & = & x_1 Q_l^2 = Q_h^2\ , \qquad\qquad
\hat{y_l} = 1-(1-y_l)/x_1 = \frac{y_h}{x_1}\ , \qquad\qquad \hat{s} = x_1 s\ ,
\nonumber\\
\hat{x}_m & = & x_h = \frac{\hat{Q_l^2}}{\hat{Q_l^2}+W_h^2}
=\frac{x_m x_1}{x_1+(1-x_1)(1-x_m)} 
\label{varrel}
\ .
\end{eqnarray}

\begin{table}[t]
\caption[]{Corrections in per cents due to initial-state radiation 
(parameters see text). 
}
\begin{center}
\begin{tabular}{|c|c|c|c|c|c|c|c|c|}
\hline
\multicolumn{9}{|l|}{$Q_l^2/$GeV$^2$}\\ \hline
$10^4\, c$ & \multicolumn{7}{c|}{} & $3.3$
\\ \hline
$10^4    $ & \multicolumn{6}{c|}{} & $-11.6$  & $-2.1$
\\ \hline
$10^3\, c$ & \multicolumn{5}{c|}{} & $-14.6$  & $-7.3$  & $-1.9$
\\ \hline
$10^3    $ & \multicolumn{4}{c|}{} &
$-15.6$  & $-8.4$  & $-4.8$  & $-1.1$
\\ \hline
$10^2\, c$ & \multicolumn{3}{c|}{} & 
$-15.9$ & $-8.6$ & $-5.3$ & $-3.4$ & $-0.6$ 
\\ \hline
$10^2    $ & \multicolumn{2}{c|}{} & 
$-16.0$ & $-8.7$ & $-5.3$ & $-3.8$ & $-2.6$ & $-0.3$ 
\\ \hline
$10^1\, c$ & \multicolumn{1}{c|}{} & 
$-16.1$ & $-8.9$ & $-5.4$ & $-3.6$ & $-2.9$ & $-2.0$ & $-0.4$ 
\\ \hline
$10$ & 
$-17.3$ & $-8.5$ & $-3.7$ & $-0.3$ & $2.0$ & $3.6$ & $5.2$ & $6.5$
\\ \hline
$c/ 10^{-4}$ 
& $c\, 10^{-4}$ & $10^{-3}$ & $c\, 10^{-3}$ & 
$10^{-2}$ & $c\, 10^{-2}$ & $10^{-1}$ & $c\, 10^{-1}$ & $1$
\\ \hline
\multicolumn{9}{|c|}{$x_m \qquad (c = \sqrt{10} \approx 3.16)$}
\\ \hline
\end{tabular}
\end{center}
\end{table}
We find the following result for the Compton 
contribution to the total cross section
\begin{equation}
  \sigma[ee \rightarrow ee X] = 
     \frac{y_l\, \rd y_l}{1-y_l}\,
     \frac{\rd x_m}{x_m\, (1-x_m)}\,
     \frac{\rd Q_l^2}{Q_l^4}\,
     \frac{\rd z}{z}\,
     \frac{\rd Q_h^2}{Q_h^2}\,
      \frac{ (1-x_m)\, Q_l^2}{ (1-x_m)\, Q_l^2 + x_m\, Q_h^2 }\,
    \Sigma
\ ,
\label{newCompton}
\end{equation}
where
\begin{eqnarray}
\lefteqn{
  \Sigma(x_h,x_l,y_l,Q^2_h,Q_l^2,z) = 
      \alpha^3\, Y_+(y_l)\, z f_{\gamma/ \re}(z) }
\nonumber \\ 
&&~ \left\{ 
  \left[ 1 + \left( 1 - \frac{x_l}{x_h}\right)^2 \right]\, F_2(x_h,Q_h^2)
  - \left( \frac{x_l}{x_h}\right)^2 \, F_L(x_h,Q_h^2) \right\}
\ .
\label{Sigmadef}
\end{eqnarray}
Note that $\Sigma$ is a very smooth function, hardly dependent on its 
arguments. Only at very low $Q_h^2$, gauge invariance forces $F_2$ (and 
hence $\Sigma$) to vanish linearly with $Q_h^2 \rightarrow 0$. The fall-off
at $x_h \rightarrow 1$ may be very slow due to the pointlike contribution to
$F_2^\gamma$ (in contrast to $F_2^p$). 

The argument $x_h$ of $F_2$ is here related to the integration variables via
\begin{equation}
  x_h = \frac{x_m\, Q_h^2}{ (1-x_m)\, Q_l^2 + x_m\, Q_h^2 }
\ ,
\label{xhxmrel}
\end{equation}
and the integration limits read 
\begin{eqnarray}
 \frac{1-x_m}{x_m}\, Q_l^2\, \frac{Q_l^2}{y_l\, z\, s - Q_l^2}
 & < &  Q_h^2 < 
  y_l\, z\, s - \frac{1-x_m}{x_m}\, Q_l^2
\nonumber\\
  {\rm Max} \left\{ \frac{\Wminsq + Q_l^2}{y_l\, s}, 
     \frac{Q_l^2}{y_l\, x_m\, s} \right\} & < &  z < 1
\ .
\label{zlowdef}
\end{eqnarray}

\begin{table}[tbp]
\caption{Corrections in per {\bf mille} due to the Compton process
(parameters see text).}
\begin{center}
\begin{tabular}{|c|c|c|c|c|c|c|c|c|}
\hline
\multicolumn{9}{|l|}{$Q_l^2/$GeV$^2$}\\ \hline
$10^4\, c$ & \multicolumn{7}{c|}{} & $15.4$
\\ \hline
$10^4    $ & \multicolumn{6}{c|}{} & $0.07$  & $1.2$
\\ \hline
$10^3\, c$ & \multicolumn{5}{c|}{} & $0.07$  & $0.3$  & $1.0$
\\ \hline
$10^3    $ & \multicolumn{4}{c|}{} &
$0.1$  & $0.4$  & $0.5$  & $1.1$
\\ \hline
$10^2\, c$ & \multicolumn{3}{c|}{} & 
$0.2$ & $0.7$ & $0.9$ & $0.8$ & $1.1$ 
\\ \hline
$10^2    $ & \multicolumn{2}{c|}{} & 
$0.2$ & $0.9$ & $1.4$ & $1.3$ & $1.0$ & $1.1$ 
\\ \hline
$10^1\, c$ & \multicolumn{1}{c|}{} & 
$0.3$ & $1.1$ & $1.7$ & $2.0$ & $1.7$ & $1.1$ & $1.1$ 
\\ \hline
$10$ & 
$0.5$ & $1.9$ & $3.3$ & $4.4$ & $4.7$ & $3.8$ & $2.6$ & $2.1$
\\ \hline
$c/10^{-4}$
 & $c\, 10^{-4}$ & $10^{-3}$ & $c\, 10^{-3}$ & 
$10^{-2}$ & $c\, 10^{-2}$ & $10^{-1}$ & $c\, 10^{-1}$ & $1$
\\ \hline
\multicolumn{9}{|c|}{$x_m \qquad (c = \sqrt{10} \approx 3.16)$}
\\ \hline
\end{tabular}
\end{center}
\end{table}
In tables~1 and~2 we present the size of the radiative corrections 
for (logarithmically distributed) bins 
ranging from $10^{-4}$ up to 1 in $x_m$ and $3.2\,$GeV$^2$ up to 
$3.2 \times 10^4\,$GeV$^2$ in $Q^2_l$. For example, the bin in the lower left 
corner corresponds to $10^{-4} < x_m < 3.2\times 10^{-4}$ and 
$3.2 < Q_l^2 < 10\,$GeV$^2$. The numbers are for a typical LEP kinematics, 
namely $2\, E_b = \sqrt{s} = 175\,$GeV, 
$W_h > 2\,$GeV, anti-tag angle $\theta_{\rm{max}} = 30\,$mrad, 
minimum tagging angle $\theta_{\rm{tag}}=30\,$mrad, 
minimum tagging energy $E_{\rm{tag}}=0.5\, E_b$, and we have used the
SaS 1D distribution functions of the photon \cite{SaS}. 
While the corrections from the Compton process are small, the correction 
from initial-state radiation are sizeable and cannot be neglected.

As an example of the distortion of the Born (non-radiative) kinematics
we show in Fig.~3 the distributions in $Q_l^2$ and $Q_h^2$ for the 
same kinematical situation: the 
scale entering the structure function ($Q_h$) does differ substantially 
from the one ($Q_l$) measured from the scattered electron.
A Fortran program (``RADEG'') 
that computes the corresponding correction factors 
either for fixed $x,y,Q^2$ or for user-defined bins in these variables, with
integration inside the bin, is available from the authors.

\clearpage

\end{document}